\documentstyle[preprint,aps,aps12,prb,epsf,epsfig,amssymb]{revtex}

\begin{document}
\relax

\begin{titlepage}
\title{Theoretical and Experimental Adsorption Studies \\ of Polyelectrolytes
on an Oppositely Charged Surface}

\author{R. Jay Mashl\footnote{Present address:  NCSA, University of Illinois, 405 N. Mathews Ave., Urbana, IL  61801}  and Niels Gr{\o}nbech-Jensen }
\address{Theoretical Division and Center for Nonlinear Studies,\\
Los Alamos National Laboratory, Los Alamos, New Mexico~~87545 }

\author{M. R. Fitzsimmons and M. L{\"u}tt }
\address{Manuel Lujan Jr.~Neutron Scattering Center,\\
Los Alamos National Laboratory, Los Alamos, New Mexico~~87545 }

\author{DeQuan Li }
\address{Chemical Science and Technology Division,\\
Los Alamos National Laboratory, Los Alamos, New Mexico~~87545 }
\vspace*{12pt}

\date{Final version published in {\it The Journal of Chemical Physics},
{\bf 110}, 2219--2225 (1999)}

\maketitle

\begin{abstract}
\setlength{\baselineskip}{0.5\baselineskip}
 Using self-assembly techniques, x-ray reflectivity measurements, and
computer simulations, we study the effective interaction between
charged polymer rods and surfaces.  Long-time Brownian dynamics
simulations are used to measure the effective adhesion force acting on
the rods in a model consisting of a planar array of uniformly
positively charged, stiff rods and a negatively charged planar
substrate in the presence of explicit monovalent counterions and added
monovalent salt ions in a continuous, isotropic dielectric medium.
This electrostatic model predicts an attractive polymer-surface
adhesion force that is weakly dependent on the bulk salt concentration
and that shows fair agreement with a Debye-H\"uckel approximation for
the macroion interaction at salt concentrations near 0.1 M.
Complementary x-ray reflectivity experiments on poly(diallyldimethyl
ammonium) chloride (PDDA) monolayer films on the native oxide of
silicon show that monolayer structure, electron density, and surface
roughness are likewise independent of the bulk ionic strength of the
solution.
\end{abstract}

\end{titlepage}

\setlength{\baselineskip}{0.5\baselineskip}

\section*{Introduction}

Coulombic interactions are ubiquitous in biological systems, as Nature
uses them in aqueous environments to regulate the structure of
biological macroions such as proteins so that desired catalytic
properties can be maintained.\cite{smilda,spassov} Whereas
electrostatic interactions are very important to biological systems,
water-soluble synthetic polymers (i.e., polyelectrolytes) also use
electrostatic interactions to gain solubility in hydrophilic environments
and, like proteins, are also expected to optimize their structures
using secondary interactions such as dipole-dipole and
hydrogen-bonding interactions.  Coulombic interactions
apply more generally to many industrial applications; for example,
they are key to controlling the stability and flocculation properties
of colloids.\cite{vanduij,see,peng}  A clear example of the
significance of electrostatic interactions is the biologically inspired
concept of molecular self-assembly on surfaces. In essence, molecular
self-assembly is a phenomenon in which hierarchical organization or
ordering is spontaneously established in a complex system without
external intervention.  Electrostatic interactions have been
successfully employed in the fabrication of layered molecular
assemblies,\cite{decher} including functional multilayered devices
such as light-emitting thin films and diodes.\cite{tfilm,fou}  Here we
discuss electrostatic interactions as a driving force in the
spontaneous self-assembly of rigid polyelectrolytes on surfaces.

Numerous analytical, simulational, and experimental polyelectrolyte
adsorption studies (for reviews, see Refs.
\onlinecite{coh88,coh91,fle93,hay94}) have examined how the amount of
adsorbed polymer and thickness of the absorbed layer depend on
properties such as the solution ionic strength, solution pH, molecular
weight or length of polymer, bulk polymer concentration, linear charge
density of the polymer, and surface potential or surface charge
density.  Nearly all of these studies focus on flexible, ``weak''
polyelectrolytes of variable degree of dissociation along the chain.
The difficulty in treating polyelectrolyte adsorption theoretically
lies in the complex interplay among chain conformational entropy and
long-ranged electrostatics.\cite{coh91,oos71,deg79,odijk79,dob95} The
entropy introduced into the system by the flexible backbones competes
with the inherent bare attraction between the oppositely charged
chains and surface.\cite{muthu2} Here, we study just one aspect of
polyelectrolyte adsorption problem: the effect of ionic strength on
the effective forces between charged
polymers and an oppositely charged surface.  Added salt has been 
noted for its dual effect on adsorption; whether or not
increasing the salt concentration leads to an increase or decrease in
the adsorbed amount depends on the balance between the screening of
intrachain repulsion and chain-surface attraction.\cite{steeg} For
hydrophobically modified polyelectrolytes, added salt can act as
a switch for adsorption.\cite{tirrell} To eliminate the issue of the
internal degrees of freedom of the chains, we treat the chains as
rigid rods so that the roles of electrostatics and ion entropy can be
studied.

\begin{figure}
\begin{center}
\epsfig{file=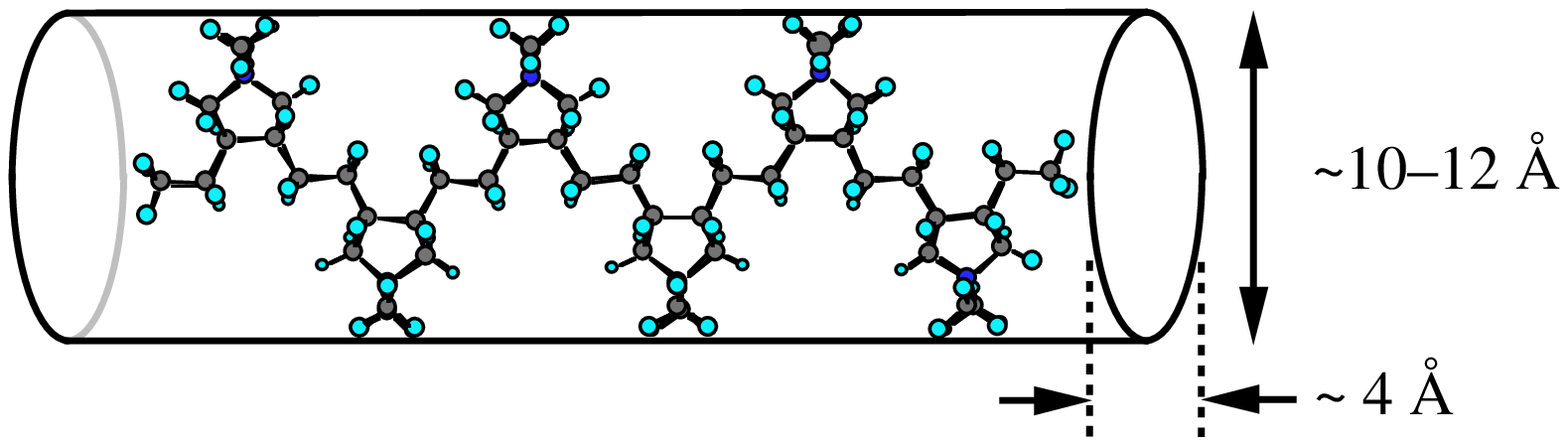,width=3.5in}
\end{center}
\caption{
  Top view of an idealized structure of poly(diallyldimethyl ammonium)
  chloride (PDDA).  Approximate dimensions of the cross section of
  the polymer vary from about 4 {\AA} to about 10--12 {\AA}.
  }\label{pdda}
\end{figure}

The most often studied self-assembling experimental systems are
self-assembled monolayers because they can be conveniently manipulated
and studied on substrate surfaces.  In particular, oxide surfaces with
their low negative charge density can be used to anchor
polycations---rather than monovalent cations---because the number of
charge-charge attractions is greater.  Since the Coulombic energy to
separate a monovalent ion pair in water initially at a distance of 5~{\AA} is
3.5 kJ/mol, or about 1.4 times the thermal energy, a single
charge-charge interaction is usually not strong enough to produce
well-organized monolayer structures.  Multiple charge attractions,
however, are able to generate good surface adhesion between films and
substrates.  Poly(diallyldimethyl ammonium) chloride (PDDA), whose
idealized structure is shown in Fig.~\ref{pdda}, is an ideal example
for this study for several reasons.  PDDA is a ``strong''
polyelectrolyte; 
its backbone charge density (and hence morphology) is not influenced by
the pH of the surrounding solution.  As it also lacks lone
electron pairs and empty orbitals, it neither participates in hydrogen
bonding nor functions as a ligand to metal ions.
Thus, the dominant 
interactions involving PDDA are expected to be 
electrostatic in nature.

In this paper we calculate the effective interaction between an array
of model rigid rods parallel to an oppositely charged interface in the
presence and absence of added salt using Brownian dynamics simulations
with explicit ions in a continuous aqueous medium.  The
resulting effective interaction is an attractive 
adhesion force that depends relatively weakly on the bulk
concentration of monovalent salt.  We also present complementary x-ray
reflectivity measurements on single-layer PDDA films on the native
oxide of a silicon substrate that show a weak dependence on ionic
strength (due to monovalent salt) in the monolayer structural
properties of electron density, surface roughness, and thickness.  The
model suggests that PDDA monolayers self-assemble via a largely
salt-independent, adhesion-attraction force when the intermolecular
interactions are governed by electrostatics.

\section*{Theory}
\subsection{Model}

\begin{figure}[H]
\begin{center}
\epsfig{file=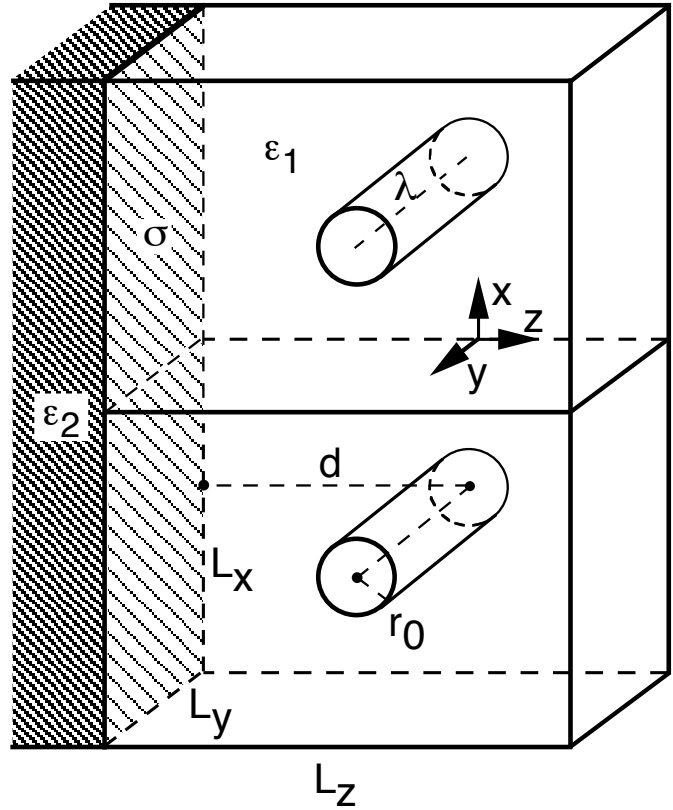,width=3.5in}
\end{center}
\caption{
  Two adjacent unit cells of sizes $L_x$ $\times$ $L_y$ $\times$ $L_z$ in the
  simulational system.  Periodic boundary conditions are applied in
  the $x$ and $y$ directions.  The line charges have a uniform charge
  density $\lambda$ and radius $r_0$ and are located a distance $d$
  from a charged surface with average charge density $\sigma$.  An
  (optional) uniformly neutral surface is located at $z=L_z$.  The
  dielectric constants for the regions with and without the line
  charges are $\epsilon_1$ and $\epsilon_2$, respectively.
  }\label{unitcell}
\end{figure}

The model consists of a combination of mobile ions and fixed macroions
together in a unit cell.  Figure \ref{unitcell} shows two adjacent
unit cells, each of which is rectangular region of dimensions
$L_x$ $\times$ $L_y$ $\times$ $L_z$ and contains a single line charge (i.e.,
PDDA) of uniform charge density $\lambda$ located a distance $d$ from
a fixed, flat surface (i.e., silicon) of a specified uniform charge
density $\sigma$ located at $z=0$.  To approximate an infinite
thermodynamic system, the unit cell is replicated in the $x$ and $y$
directions using periodic boundary conditions to produce an infinite one-dimensional array of infinitely
long, parallel line charges of spacing $L_x$ and repeat distance $L_y$
in the $y$ direction, parallel to a charged, infinite plane. 
The $z$ direction is not periodic.  The dielectric
constant within the unit cell is $\epsilon_1$, and that of the medium
below ($z \leq 0$) is $\epsilon_2$.  For simplicity, the case 
where $\epsilon_1 = \epsilon_2$ is studied, thereby producing no dielectric
interface.  Counterions and co-ions, consisting of monovalent cations
and anions, are added such that the system is overall charge neutral.
To study systems having a ``bulk'' concentration of salt, a uniformly
charge-neutral surface is placed at $z=L_z$ to confine the
particles during the simulation.

As exact correspondences between the physical parameters
describing real PDDA polymers and silicon surfaces with a native oxide
layer are difficult to
make, we make the following approximations.  
Real PDDA polymers (Fig.~\ref{pdda}) have one
positive charge per 5.4 {\AA} and have an anisotropic diameter, ranging
from 4 to 12 {\AA} due to their molecular structure.  The model rod
is chosen to have a uniform axial linear charge density $\lambda =
e/10$ {\AA} and a radius $r_0$ of an intermediate value of 4~{\AA}.
This rod size enters through a short-ranged repulsion between the
rod axis and the mobile ions and is discussed in the next section.  The rod
spacing $L_x$ is taken to be 40 {\AA}.  The native oxide on silicon
and its distribution of negatively charged hydroxyl groups is
represented by a uniform surface charge density $\sigma =
-e/60$ {{\AA}}$^{2}$.  The dielectric constants of the aqueous medium
and substrate interior were both set equal to 80, and all simulations
were carried out at room temperature.

Each simulation further required the specification of the values for
the rod-center-to-surface separation distance $d$, the repeat rod
segment length $L_y$, unit cell height $L_z$, and the numbers and
charges of the mobile particles.  The values of $L_y$ and $L_z$ fell
into the ranges 60--150 {\AA} and 60--120 {{\AA}}, respectively, and
approximately 100 ions were introduced at random into the unit cell
until the system became charge neutral.  The Brownian dynamics
algorithm \cite{allen} used to simulate the motion of the ions relates
the positions ${\bf r}_i $ of the ions $i$ at a time $t+\Delta t$ to those
at the previous time $t$ according to the relation ${\bf r}_i(t+\Delta
t) = {\bf r}_i(t) + (D\Delta t/ k_{\rm B}T){\bf f}_i(t) + {\bf
  r}^*_i(\Delta t)$, where ${\bf f}_{i}(t)$ is the deterministic force
acting on ion $i$ due to long-ranged electrostatic and
short-ranged nonelectrostatic interactions, ${\bf r}^*_{i}(\Delta t)$
represents the random displacement of ion $i$ due to the random
thermal motions of a discrete solvent, $D$ is the isotropic diffusion
constant, $k_{\rm B}$ is Boltzmann's constant, and $T$ is temperature.
This description assumes the solvent to be a continuum.  The diffusion
constant $D$ is related to the particle mass $m$ and the coefficient
of friction $\xi$, due to a particle moving against the solvent, by
the relation $D=k_{\rm B}T/m\xi$.  In all simulations $m\xi\equiv 1$
and the time step $\Delta t$ was 0.005, in normalized units.  At every
time step the force ${\bf f}_{i}(t)$ is
calculated as the gradient of the potential energy surface due to the
ions, rods, and the surface, and the random displacement is chosen
independently for each particle from a Gaussian distribution with a
variance of $2k_{\rm B}T\Delta t$ in each spatial
component.\cite{parisi} The positions of the particles were updated
according to the periodic boundary conditions.  During the simulations
the average vertical $z$-force on the rod (i.e., the component of the
force acting on the rod in the direction perpendicular to the charged
surface) and the distributions of the ions were monitored. The system
was considered to have reached equilibrium when the time-averaged
vertical force on the rod reached a steady value, the average lateral
$x$-force on the rod was zero, the total system energy reached a
steady value, and the ion distributions remained stationary.  These
requirements necessitated the time averages to be accumulated for up
to 11 $\times$ 10$^6$ time steps after the initial 5--25 $\times$ 10$^4$
steps were discarded.

\subsection{Interaction potentials}

The interaction potentials used in the simulations consist of
pairwise, long-ranged electrostatic forces and
short-ranged, nonelectrostatic repulsions, where the electrostatic
interactions are exact for these periodic systems.  The ion-surface
electrostatic interaction $V_{\rm is}(z)$ per unit cell is a function
of the distance $z$ between an ion with charge $q$ and the surface: 
\FL\begin{eqnarray}
\quad\quad V_{\rm is}(z) = -{q \sigma \over 2 \epsilon_1\epsilon_0} z,
\end{eqnarray}
where $\epsilon_0$ is the vacuum permittivity.  The rod-surface
electrostatic contribution $V_{\rm rs}(z)$ per
unit cell for a uniform surface charge distribution is similarly 
\FL\begin{eqnarray}
\quad\quad V_{\rm rs}(z) = -{\lambda L_y \sigma \over 2 \epsilon_1\epsilon_0} z.
\end{eqnarray}
The ion-ion electrostatic potential energy $V_{\rm ii}(\Delta{\bf
r})$ per unit cell 
for an ion with charge $q_1$ at the point $(x + \Delta x, y +
\Delta y, z + \Delta z)$ in the unit cell 
and an ion with charge $q_2$ in the unit cell and its
replicas located at $(x+mL_x,y+nL_y,z)$, where $m,n$ are integers,
resulting from the two-dimensional replication of the unit cell in
the $x$ and $y$ directions, is \cite{ngj_1oR} 
\FL
\begin{eqnarray}
\quad\quad V_{\rm ii}(\Delta{\bf r}) = {q_1 q_2\over 
4\pi\epsilon_1\epsilon_0}\Biggl\{ {4 \over L_x }\sum^\infty_{n=1} 
\cos\left(2\pi{\Delta x\over L_x}n\right) \sum^\infty_{k=-\infty} \Biggr. \nonumber 
\end{eqnarray}
\vspace{-1.5\baselineskip}
\FL\begin{eqnarray}
\qquad K_0\left\{ 
2\pi n \left[ 
\left({L_y\over L_x}\right)^2 \left({\Delta y\over L_y}+k\right)^2 +
\left({\Delta z\over L_x}\right)^2
\right]^{1/2} 
\right\} \nonumber
\end{eqnarray}
\vspace{-1.5\baselineskip}
\FL\begin{eqnarray}
\qquad  -{1\over L_x}\ln\left[\cosh\left(2\pi{\Delta z\over L_y}\right) -
\cos\left(2\pi{\Delta y\over L_y}\right)   \right] \nonumber 
\end{eqnarray}
\vspace{-1.5\baselineskip}
\FL\begin{eqnarray}
\label{eqn_vii}
\qquad \Biggl. -{\ln 2 \over L_x} \Biggr\},
\end{eqnarray}
where $K_0$ is the modified Bessel function of the second kind of
order zero.
The ion-rod potential energy $V_{\rm
ir}(\Delta{\bf r})$ per unit cell is the combined logarithmic interactions between a point
particle and a one-dimensional array of line charges.  The analytic
form of $V_{\rm ir}(\Delta{\bf r})$ is derived from the potential
energy of an ion interacting logarithmically with a two-dimensional
array of line charges arranged on a rectangular lattice (see Eq.~(14)
in Ref.~\onlinecite{ngj_logR}) by eliminating one of the dimensions.  The
result is 
\FL\begin{eqnarray}
\quad\quad V_{\rm ir}(\Delta{\bf r}) = -{q\lambda\over 4\pi\epsilon_1\epsilon_0} 
   \ln \left\{ 2 \left[ \cosh\left(2\pi{\Delta z\over L_x}\right) \right. \right. \nonumber 
\end{eqnarray}
\vspace{-1.5\baselineskip}
\FL\begin{eqnarray}
\label{logint}
\label{eqn_vir}
\qquad\qquad\qquad \left. \left. - \cos\left(2\pi{\Delta x\over L_x}\right)
\right] \right\}.
\end{eqnarray}
The associated ion-ion and rod-rod self-energies arising from an
ion/rod interacting with its own periodic replicas are given elsewhere
\cite{ngj_1oR,like_paper} and need not be considered in this work.

Finally, the ion-ion, ion-rod, ion-surface, and rod-surface
short-ranged, nonelectrostatic repulsions were modeled as $A_{\rm
  ii}/r^{12}$, $A_{\rm ir}/r^{11}$, $A_{\rm is}/r^{10}$, and $A_{\rm
  rs}L_y/r^{10}$, respectively, to prevent electrostatic collapse of
the charge-neutral system.  The combination of Coulombic attraction
and short-ranged repulsion between two oppositely charged ions, ion
and rod, ion and surface, or rod and surface introduces optimal
ion-ion, ion-rod, ion-surface, and rod-surface distances,
respectively.  Chemically speaking, these optimal distances are a
measure of the ``polar-bond'' distance between oppositely charged
species.  Values of the $A$ coefficients were $A_{\rm ii} = 5.26$
$\times$ 10$^3$ kcal {\AA}$^{12}$/mol, $A_{\rm ir} = 3.17$ $\times$
10$^5$ kcal {\AA}$^{11}$/mol, $A_{\rm is} = 6.61$ $\times$ 10$^2$ kcal
{\AA}$^{10}$/mol, and $A_{\rm rs} = 1.82$ $\times$ 10$^4$ kcal
{\AA}$^{9}$/mol, giving optimal ion-ion, ion-rod, ion-surface, and
rod-surface distances of 2.4, 4.0, 2.4, and 4.0 {\AA}, respectively.
In general, in order to use a line-charge model to represent a PDDA
polymer and its nonaxially distributed ammonium charge centers, the
model rod radius $r_0$ differs from the real size of the polymer so
that the electrostatic ``binding'' energy between a counterion and a
PDDA charge---on the order of a few $k_{\rm B}T$---can be obtained.
The ion-rod short-ranged interaction was applied to all mobile charges
and was taken according to the minimum image convention.\cite{allen}

\section*{Experimental}

The preparation of self-assembled PDDA monolayers has been described
previously.\cite{lutt} Here we summarize the main points and describe
the differences from previous experiments.  The growth of PDDA
monolayers was carried out on thin silicon wafers instead of thick
silicon substrates.  The PDDA solution concentration used for these
experiments was 0.1 M instead of 1 mM, and the ionic strength of these
solutions was tuned with monovalent salt (NaCl) to obtain ionic
strengths of $I$ = 0.001, 0.01, and 0.1 M.  The reaction time for
deposition of PDDA onto the substrate was extended from 5 min to
20 min at room temperature.  The x-ray reflectivity measurements were
carried out as described previously without modification, and the
quality of the data for thin silicon wafers (500 $\mu$m) is the same
as those of thick silicon substrates (0.1 cm).

\section*{Results and Discussion}

\begin{figure}[H]
\begin{center}
\epsfig{file=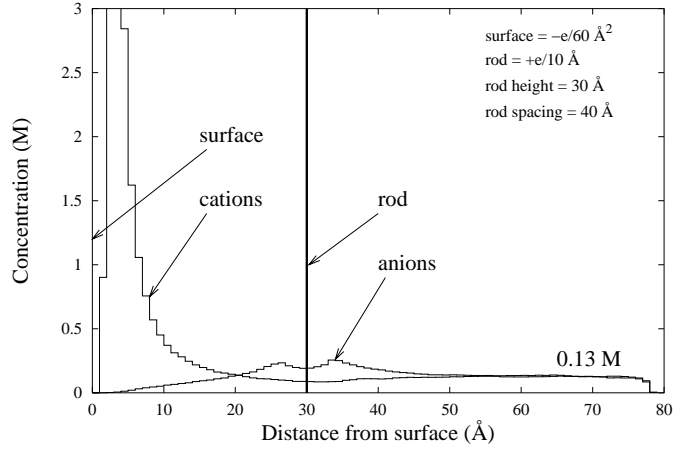,width=3.5in}
\end{center}
\caption{
  Distributions of mobile ions, averaged over the $x$ and $y$ coordinates in
  the unit cell, in an equilibrated system, showing high
  concentrations of countercharge near the macroions and a plateau at
  0.13 M for distances $z \gg d$.  The unit cell size is $(L_x, L_y,
  L_z) = (40, 120, 80)$ {\AA}.
}\label{distribs}
\end{figure}

To see the qualitative similarity between theory and experiment on the
structural properties of monolayers of rigid rods near oppositely
charged surfaces, we first examine the results of the particle
simulations.  Figure \ref{distribs} shows a typical concentration
profile for the mobile ion species at equilibrium.  The time-averaged
monovalent cation distribution near the surface and monovalent anionic
distribution near the rod are noticeably peaked.  At large distances
from both the rod and surface, the concentration profiles are
featureless.  This flat region is identified as the bulk electrolyte
solution, whose salt concentration $c_s$ is given by the height of the
plateau.  The concentration in Fig.~\ref{distribs} is approximately
0.13 M.  
Clearly, the local concentration of ions can differ significantly from
the bulk.  For the surface charge density used in the simulations,
the ion concentration near the surface is two orders of magnitudes
greater than the bulk (not shown).  The bulk concentration as
indicated by the simulation corresponds to the
concentration that would be measured experimentally.

\begin{figure}[H]
\begin{center}
\epsfig{file=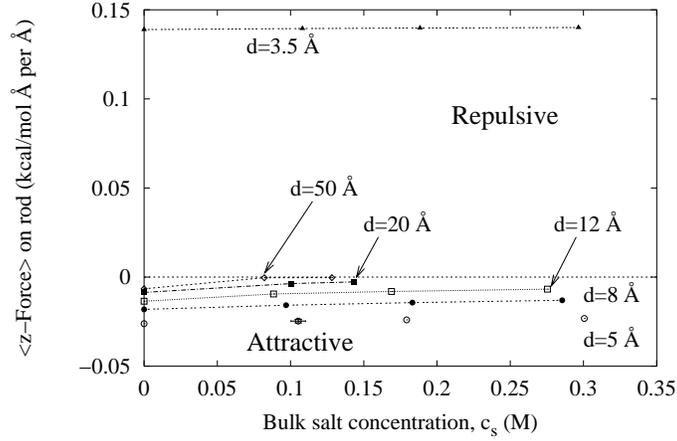,width=3.5in}
\end{center}
\caption{
  The simulated equilibrium adhesion force acting on a rod and its
  dependence on the bulk monovalent salt concentration $c_s$ for several
  rod-center-to-surface distances, $d$.  For clarity, linear
  interpolation is used to join the data, except for the case
$d=5$ {\AA} where one set of error bars is shown.
}\label{rawadhesion}
\end{figure}

\begin{figure}[H]
\begin{center}
\epsfig{file=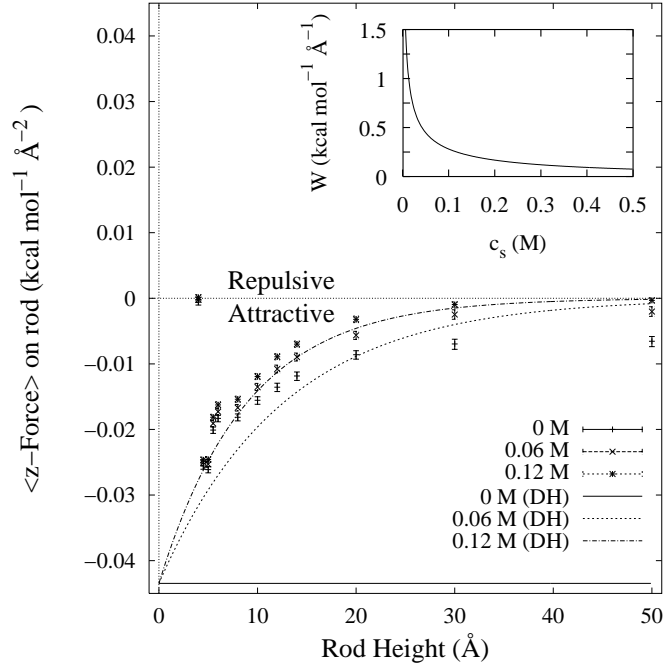,width=3.5in}
\end{center}
\caption{
  Linearly interpolated values ($+$, $\times$, and *) 
for the equilibrium adhesion force on a rod,
  derived from Fig.~\ref{rawadhesion}, as a function of
  rod-center-to-surface distance, $d$, for several values of the bulk
  electrolyte concentration $c_s$.  Also shown are the predictions of
  Debye-H\"uckel theory for macroions (lines), given by
  Eq.~(\ref{dheqn}).  Inset:  Adhesion energy per rod and its variation
with bulk salt concentration for a rod-surface equilibrium distance 
$d^*$ = 4~{\AA}.
The parameter values are
  $\sigma=-e/60$ {{\AA}}$^2$, $\lambda=e/10$ {\AA}, and
  $L_x=40$ {\AA}.
}\label{interpadhesion}
\end{figure}

The time-averaged vertical $z$-force on the rod that we call the 
adhesion
force and its dependence on $c_s$ 
for sample values of the fixed rod-surface distance $d$ is
shown in Fig.~\ref{rawadhesion}.  Force values greater than zero
indicate an effective rod-surface repulsion; less than zero, an
attraction.  Here, a bulk concentration of zero means that only a
charge-neutralizing number of counterions was added to the simulation
unit cell with the neutral lid removed.
The data reveal that the effective attraction decreases
rather weakly with increasing salt concentration 
for all distances $d$.  For small $d$, this behavior may
simply be
the result of salt exclusion from between the rod and the surface.
An alternate way of viewing the effect of $c_s$
on the adhesion force is shown in
Fig.~\ref{interpadhesion}, where the force ($+$, $\times$, and
*) is now plotted against the rod-surface spacing $d$ for
several values of the bulk salt concentration.
Figure \ref{interpadhesion}, obtained by linear interpolation of the
data sets shown in the previous figure, shows three major features:
1) the force at $d=4$ {\AA} is approximately zero; 2) there is a
shoulder in the force curve near $d\approx 6$ to 8 {\AA}; and 3) the
effective attractive force generally decreases with increasing
rod-surface distance, regardless of $c_s$.  The
fact that the force crosses over from attraction to repulsion near
$d=4$ {\AA} results from our having fixed the equilibrium rod-surface
distance in the absence of any ions at 4 {\AA} in the model via the
short-range repulsive coefficients discussed earlier.  The shoulder is
likely due to the finite size of the ions because the ions are
expected to be able to pass freely in between the rod and the surface
only for $d > 6.4$ {\AA}.

The interpolated force-distance data was compared to Debye-H\"uckel
theory\cite{mcquarrie} (DHT), adapted to interactions between
macroions.  Briefly, DHT is a mean-field theory that gives an
exponentially screened Coulombic electrostatic interaction between two
ions due to their
ionic atmospheres.  The
adhesion force per unit length of rod acting on each rod as a function
of the rod-surface distance $d$ is
\FL
\begin{equation}
\quad\quad {\bf f}_{\rm DHT}(d) =
{\sigma\lambda\over2\epsilon\epsilon_0} e^{-\kappa d}\ {\bf\hat{z}},
\label{dheqn}
\end{equation}
where $\kappa^{-1}$ is the Debye 
screening length, given in terms of the (bulk) ionic
strength $I$ as $\kappa^2 = 2Ie^2/\epsilon\epsilon_0 k_{\rm B} T$.  
  As a guideline, DHT generally
provides an adequate description of electrostatic interactions between
two ions when their interaction energy is small compared to $k_{\rm B}T$.
The above expression for the 
adhesion force, if it is valid, is thus expected to provide better 
predictions for the adhesion force 
as $\kappa d$ increases.
The results of DHT
are shown in Fig.~\ref{interpadhesion} as the solid and broken lines
for bulk ionic strengths of 0, 0.06, and 0.12 M, corresponding to
$\kappa$ values of 0, 0.08, and 0.11 {\AA}$^{-1}$, respectively,
as derived from the simulation data.  At
small rod-surface separation distances, agreement between DHT and the
simulations not surprisingly fails because the simulations include a
short-ranged rod-surface nonelectrostatic repulsion that is absent from DHT.  Under
conditions of zero ionic strength, the simulations and DHT disagree
severely as a result of the incompatibility of DHT and our method for
determining $c_s$ from the simulations.  
Although DHT can in principle account for screening due to counterions
alone, the formulation of ${\bf
  f}_{\rm DHT}$ implicitly assumes that screening is primarily due
to added salt.  In the absence of salt, the 
resulting electrostatic interaction is that of 
uniformly charged, bare macroions.  At ionic strength
values of 0.06 and 0.12 M, DHT is seen to capture fairly
well the behavior of the interpolated force-distance curves
for distances beyond the shoulder ($d > 8$ {\AA}), with better
agreement occurring for the larger ionic strength values.
Interestingly, the simulations and DHT results seem comparable for $d
> 5$ {\AA}, although the validity criterion on the electrostatic
interaction energy is 
marginally satisfied for the range of rod-surface
distances shown.  Finally, there may still be a
qualitative disagreement in the force between the simulations and DHT for $d >
50$ {\AA}.  Whereas DHT gives an adhesion force 
on the rod that decays exponentially with distance, simulations
indicate a more slowly decaying force and may be 
due to the increased equilibration
times needed for large rod-surface separations.

Under the condition that DHT in Eq.~(\ref{dheqn}) is a good approximation
to the adhesion force curve, the adhesion or
``dissociation'' energy per unit length rod per rod may be calculated as
the integral of the force:
\FL
\begin{equation}
\quad\quad W(d^*) = \int_\infty^{d^*} dz\ f_{\rm DHT}(z) =
-{\sigma\lambda\over 2\epsilon_1\epsilon_0\kappa} e^{-\kappa d^*},
\end{equation}
where $d^*$ is the equilibrium distance of the rod from the surface.
The inset of Fig.~\ref{interpadhesion} shows the adhesion
energy per rod as a strongly decaying function of the bulk salt concentration
$c_s$ that goes as $W(d^*) \sim c_s^{-1/2} \exp(-c_s^{1/2}d^*)$. The
DHT model gives a substantial adhesion energy: about 30 kcal/mol for 
every 10 nm of rod in a 0.1 M (monovalent) salt solution.

The theoretical model suggests that in experiments, where the polymer
rods are mobile in solution, the rods would be attracted to the
surface by a fairly salt-independent adhesion force and thus would move
spontaneously toward the surface and possibly form a monolayer.
Indeed, monolayer formation is observed in the molecular self-assembly
of PDDA polymer as evidenced by x-ray reflectivity measurements.  PDDA
was found to form a uniform nanometer-thick thin film on a silicon
substrate at various solution ionic strengths in the range 
0.001 to 0.1 M.  Other film properties, such as the film electron
density and surface roughness, were also found to be independent of
the bulk salt concentration.  The driving force for the formation of
the monolayers on the surface is the electrostatic attraction between
the charged rods and surface.

\begin{figure}[H]
\begin{center}
\epsfig{file=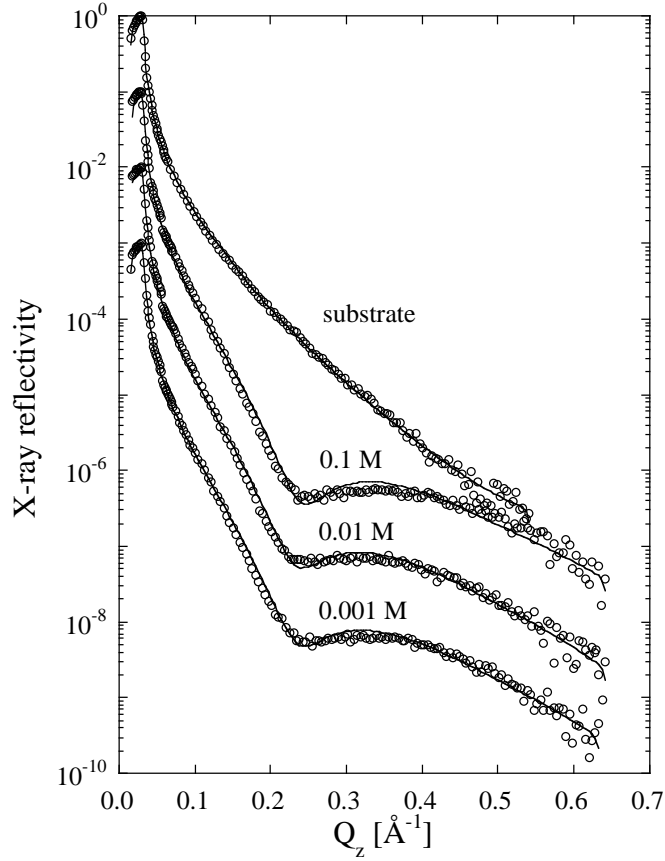,width=3.5in}
\end{center}
\caption{ Reflectivity profiles $R(Q_z)$, given by the open circles, of the
  silicon wafer terminated by its native oxide after being subjected
  to PDDA solutions of varying ionic strengths ($I$ = 0.001, 0.01,
  0.1 M NaCl).  The calculated reflectivity profiles (solid lines) are
  a fit to a model\cite{parratt} of a single film.
}\label{xray}
\end{figure}

To verify the weak influence of ionic strength on the surface and
PDDA rod adhesion characteristics, we performed x-ray reflectivity
characterization of the monolayers by measuring their reflectivity
profiles.  The reflectivity profiles $R(Q_z)$, normalized to unit
reflectivity, are shown as a function of momentum transfer $Q_z$ in
Fig.~\ref{xray}. The maximum of the reflectivity profile occurs at a
value of $Q_z$ corresponding to the condition when the x-ray radiation
in the sample is evanescent ($Q_z < Q_c$), \cite{parratt} and the
sample surface subtends the full width of the x-ray beam. A
difference, or contrast, between the electron densities of the PDDA
monolayer and the silicon substrate produces fringes and oscillations
in the x-ray reflectivity.  The amplitude of the fringes is related to
the magnitude of the contrast in electron densities.  The
oscillations with $Q_z$ are caused by interference between the x-ray
beam reflected by the film-air and film-substrate interfaces, and the
period of the oscillation is inversely related to the film
thickness. In addition to the decay in the reflectivity of the sample
with $Q_z$ due to the Fresnel reflectivity, the reflectivity profile
may be further attenuated by roughness at the interfaces.  This decay
is related to the variation in the displacement of the interface in
the direction normal to the surface about a mean value across the
sample.  The fluctuation in interface height forms a distribution
whose root-mean-square width, $\sigma_i$, increases the attenuation of
the reflectivity profile with $Q_z$.

The x-ray reflectivity data was fitted to a model \cite{parratt} for
single-layer films on a substrate which yields the average electron
density of the film, $\rho_e$ , the thickness of the film, $\Delta$,
and the surface roughness of the film, $\sigma_r$.  The values of
these parameters were determined for the PDDA monolayers by perturbing
the values from initial guesses until the weighted difference between
the observed data ($\circ$ in Fig.~\ref{xray}) and the fitted
profile was minimized, and the resulting calculated reflectivity
profiles are shown as the solid curves in Fig.~\ref{xray}.  For PDDA
monolayers formed from solutions of ionic strengths of 0.001, 0.01,
and 0.1 M NaCl, the electron density values were calculated to be,
respectively, $\rho_e = 0.225$, 0.225, and 0.234 $e^-/{\rm{\AA}}^3$,
with thicknesses $\Delta$ of 12.7, 12.8, and 12.4 {\AA}, and surface
roughnesses $\sigma_r$ of 1.2, 1.2, and 0.9 {\AA} for the PDDA-air
interfaces.  Not only are these thickness values consistent with the
formation of a monolayer of PDDA whose molecules are upright, as shown
in Fig.~\ref{pdda}, but also the fitted profiles reveal that the
monolayer structural parameters are independent of the bulk ionic
strength of the initial PDDA solution.  Thus, these experimental
results support the theoretical simulation model in that ionic
strength, as varied from 0.001 to 0.1 M experimentally and from 0 to
0.12 M theoretically, does not greatly affect the adhesion
characteristics of the polymer rods to substrate surfaces.

\section*{Conclusions}

We have developed a theoretical simulation model to predict
self-assembly behavior of rod-like polymers in aqueous salt solutions
based on electrostatic interactions and presented supporting experimental
x-ray reflectivity data for PDDA monolayers.  The model shows that the
bulk ionic strength, due to monovalent salt, has only a small effect
on the effective attractive force between model rods and an oppositely
charged surface.  Experimental x-ray reflectivity results demonstrated
that PDDA monolayer structure (thickness) and morphology (roughness)
do not vary significantly over the two orders of magnitude of 
solution ionic strength studied.
 Comparison of the model results to the prediction of
Debye-H\"uckel theory (DHT) for macroions revealed fair agreement,
even though the expected range of validity of DHT likely lies outside
the region of parameter space studied by the simulations, and the
reason for this apparent agreement is not well understood.  
Nonetheless, the theoretical and complementary experimental results of
the adhesion of charged rigid rods onto an oppositely charged
 substrate are in good agreement.

Dynamic simulations involving mobile rods and ions would be
interesting, as the kinetics\cite{van94,hoo96,sch96,stu97,fil98,pag98}
of the adsorption process could be studied.  Whereas we have shown in
this paper that a planar array of rods is attracted to the surface via
electrostatic  interactions, new information as to 1) the surface
distribution of the rods and 2) whether the rods adsorb independently
or form bundle-like structures in solution prior to adsorption could
be obtained.  However, such simulations are expected to 
depend sensitively on 
the model rod parameters, in particular, on the rod size and on
the distribution
of the charged sites.  It was recently shown\cite{us} that for two
isolated, {\it like}-charged rods in an infinite space under no-salt
conditions that the rod size is a crucial control parameter for determining
whether (divalent) counterions could mediate an effective
attraction between the rods.  This finding suggested that for a given
linear charge density of the rods, there was a maximal rod size that
would allow the rods to be mutually attracted.  Similar conclusions about
the rod size were reached in systems of like-charged rods and surfaces using
the geometry discussed in this paper.\cite{like_paper}  Nontrivial 
charge distributions on the rod surface are expected to
complicate matters further.

Whereas we have focused here on electrostatic interactions, other
interactions such as hydrogen bonding, hydrophobic effects,
and explicit dipole interactions are important in 
materials in many
fields of research.  However, the idea of incorporating combined
secondary interactions in the design of new synthetic macromolecular
materials remains largely unexplored systematically.\cite{muthu} We
anticipate that further modelling of electrostatic and other
interactions will not only provide insight into the structure of
biocompatible polymers in solution, but also lead to better design and
construction of electronic and optical devices using molecular
self-assembly techniques.

\section*{Acknowledgments}
We thank Robijn Bruinsma, William Gelbart, Li-Chung Ku, and
Philip Pincus for helpful discussions.
This work was performed under the auspices of the United States
Department of Energy, supported in part by Contract 
No.~W--7405--Eng--36 (MF, ML, and DL), by funds provided by the
University of California for conduct of discretionary research by Los
Alamos National Laboratory (NGJ), and by NSF Grant No.~DMR--9708646 (RJM).

\vspace*{0.25in}


\pagebreak

\end{document}